\documentclass{aa}
\usepackage{epsfig}
\usepackage{graphicx}
\usepackage{mathtools}
\usepackage{txfonts}
\usepackage{lscape}
\usepackage{hyperref}
\usepackage{color}
\usepackage{amsmath}
\providecommand{\sorthelp}[1]{}

\bibpunct{(}{)}{;}{a}{}{,}

\begin{document}

\title{
Detailed 3D structure of Orion\,A in dust with Gaia DR2
}
\subtitle{}

\author{Sara Rezaei Kh.\inst{\ref{inst1}{\ref{inst2}}} \and Coryn A.L. Bailer-Jones\inst{\ref{inst2}} \and Juan D. Soler\inst{\ref{inst2}} \and Eleonora Zari\inst{\ref{inst2}}\\
}
\institute{Chalmers University of Technology, Department of Space, Earth and Environment, 412 93 Gothenburg, Sweden \label{inst1} 
\and 
Max Planck Institute for Astronomy (MPIA),  K\"onigstuhl 17, 69117 Heidelberg, Germany \label{inst2} 
}

\abstract{The unprecedented astrometry from Gaia DR2 provides us with an opportunity to study in detail molecular clouds in the solar neighbourhood. Extracting the wealth of information in these data remains a challenge, however. We have further improved our Gaussian Processes-based, three-dimensional dust mapping technique to allow us to study molecular clouds in more detail. These improvements include a significantly better scaling of the computational cost with the number of stars, and taking into account distance uncertainties to individual stars.
Using Gaia DR2 astrometry together with 2MASS and WISE photometry for 30\,000 stars, we infer the distribution of dust out to 600\,pc in the direction of the Orion\,A molecular cloud. We identify a bubble-like structure in front of Orion\,A, centred at a distance of about 350 pc from the Sun. The main Orion\,A structure is visible at slightly larger distances, and we clearly see a tail extending over 100\,pc that is curved and slightly inclined to the line-of-sight. The location of our foreground structure coincides with 5-10 Myr old stellar populations, suggesting a star formation episode that predates that of the Orion Nebula Cluster itself.  We identify also the main structure of the Orion\,B molecular cloud, and in addition discover a background component to this at a distance of about 460 pc from the Sun. Finally, we associate our dust components at different distances with the plane-of-the-sky magnetic field orientation as mapped by Planck. This provides valuable information for modelling the magnetic field in 3D around star forming regions.}


\maketitle

\section{Introduction}\label{sec:intro}

The Orion molecular complex is the nearest site actively forming massive stars in the Galaxy \cite{Menten07, Bally08}. As a nearby laboratory, Orion has been studied in various aspects, ranging from distance estimates to different parts of the cloud to the star formation processes and individual stellar populations \citep[e.g.][]{Brown94, Menten07, Jeffries07, Bally08, Alves12, Bouy14, Schlafly15, Zari17,Kounkel18,Zari19}.
Stars in the Orion region are also known to be responsible for the creation of the Orion-Eridanus superbubble, a large cavity in the vicinity of the Orion that extends to the constellation of Eridanus in the sky \citep[e.g.][]{Bally08, Pon15}. Bubble structures are very common in the interstellar medium (ISM). They are results of the presence and evolution of young, massive stars that influence their surrounding ISM through radiations, stellar winds and supernovae explosions \citep[e.g.][]{Heiles79, MacLow89}.

Although the projected picture of the Orion region in the plane of the sky has been determined from various observations of the gas/dust emission \citep[e.g.][]{Ochsendorf15, Soler18}, the distance to different parts of the cloud is still debated. Orion\,A, the giant molecular filament situated in projection inside Barnard's loop, is possibly the most studied target in this vicinity. Home to the Orion Nebulae Cluster (ONC) at a distance of $\sim$ 400 pc, Orion\,A consists of rich clusters of young stars and active sites of massive star formation. \cite{Schlafly15} demonstrated that Orion\,A is not a flat filament in the plane of sky. By estimating the distance gradient along the filament, they suggested that the southern part of the filament (hereafter $\emph{tail}$) is further away from the Sun than the northern part (hereafter $\emph{head}$) hosting the ONC. Recent work by \cite{Grossscheld18} revealed an extended tail of Orion\,A to larger distances than previously estimated based on the distribution of the young stellar objects (YSOs) in the Orion\,A vicinity. \cite{Zucker19} confirmed the distance gradient for Orion\,A by estimating the distance for individual sight-lines along the filament.

One way of probing the structure of the cloud is through mapping its full three-dimensional (3D) dust distribution. Dust, only a tiny fraction of matter in the Galaxy, scatters, absorbs and re-emits light making distant objects look fainter than they are. In addition to having negative effects on observations of more distant objects, dust plays important roles in creating and shaping the ISM. It protects molecules from the high energy UV radiation, which would otherwise impede star formation. The ISM and the life cycle of stars are tightly related. Therefore studying different properties of the ISM, including the 3D distribution of dust in the Galaxy, can provide valuable information about the structure of the ISM and potential sites of star formation in the Milky Way. Furthermore, the 3D distribution of dust provides crucial information to model the distribution and dynamics of star-forming clouds. For example, knowledge of distances to various dust over-densities along a line of sight (l.o.s) is helpful to disentangle the components responsible for the plane-of-the-sky magnetic field orientations derived from submillimeter polarisation observations, such as those by ESA's {\it Planck} satellite \citep{planck2014-XXI}.

Various dust extinction mapping techniques have been developed over the past few years. While most of these methods infer the individual line-of-sight extinction towards stars, then take the derivative of these individual sight lines to get the extinction per unit distance \citep[e.g.][]{Marshall06, Sale12, Hanson14, Schlafly15, Green18}, dramatic artefacts produced in these approaches make it hard to explore the physical properties of the ISM, like the structure of the molecular clouds. \cite{Green19} tried to overcome this drawback by applying a smoothing function on the extinction derivatives using an iterative approach. \cite{Lallement19} focused on mapping the differential extinction in 3D by taking into account the neighbouring correlations. Although the approach of \cite{Lallement19} has the advantage of producing smooth maps, it does not consider distance and extinction uncertainties when inferring the differential extinction, therefore could be biased due to the data quality cuts.

Probing the 3D distribution of the dust towards the Orion region can provide valuable information about the distance to and structure of the cloud. The 3D dust mapping towards the Orion complex by \cite{Schlafly15} revealed the Orion dust ring. To further study the dust distribution towards Orion, as shown in \citep{Rezaei_Kh_17, Rezaei_Kh_18b}, we have developed a non-parametric 3D dust mapping technique that takes into account the 3D correlation between points in space, allowing the model to trace arbitrary dust variation. In \citep{Rezaei_Kh_18a} we mapped the dust distribution towards the Orion region and demonstrated the capability of our method to capture different dust clouds in the 3D space without having the discontinuities and artefacts presented in other works. The latest data release from the second release of the Gaia satellite \citep[Gaia DR2,][]{Gaia18a} with unprecedented astrometry enables us to further study this area.

In the present work we improve our mapping technique by including both distance and extinction uncertainties, together with resolving the computational limitations, thereby allowing us to exploit a large dataset like Gaia DR2 as the input, and to produce a detailed 3D dust map of the Orion\,A. We also compare the location of our dust clouds with that of young stellar populations in the region, then present an analysis of the magnetic field orientation using our density structures.


\section{3D mapping technique} \label{sec:method}

Here we briefly summarise our modelling approach (refer to \cite{Rezaei_Kh_17} and \cite{Rezaei_Kh_18b}) and also explain the improvements made.

We use a non-parametric model to infer the local dust density using the position and l.o.s attenuation to a number of stars in 3D space. The attenuation (a) is related to the extinction (A) as $A \simeq 1.0857 a$. We divide each l.o.s into small 1D cells to approximate the attenuations as the sum of the dust densities in the cells along the l.o.s to the stars. Afterwards, we connect all these cells in the 3D space using a Gaussian process that takes into consideration the neighbouring correlations between points. The closer two points are in the physical space, the more correlated their local dust densities. We use a truncated covariance function in order to account for the correlations, i.e. the points will be correlated only if they are closer than a correlation length. This way we set up our model to then infer the dust density for any arbitrary point in this space, even along a l.o.s that was not initially observed. In addition to the specification of the cell size (with fixed length), our model has three hyper-parameters: ${\lambda}$ which is the correlation length, ${\theta}$ which sets the amplitude of the density variance, and the mean of the Gaussian process. We set these hyper-parameters based on the input data \citep[using the approach described in][]{Rezaei_Kh_17, Rezaei_Kh_18b}.

As we explained in the aforementioned papers, there were a couple of limitations in our model that we overcome in the present work. 
One of the main drawbacks of the model was that we did not take into account distance uncertainty. This limited us to only using data with very precise distance measurements. Since the position of input stars and predicting points are ``given'' in our model \citep{Rezaei_Kh_17, Rezaei_Kh_18b}, any direct application of the distance uncertainty in the analytic solution is not possible. We have overcome this issue by propagating the distance uncertainty of a star into its attenuation uncertainty.
This way we have 
\begin{eqnarray}
a = \rho r \nonumber \\
{\sigma}_{{a}_{d}} = \frac{a}{r} {\sigma}_{r}
\label{eq:err} 
\end{eqnarray}
where a is the attenuation, $\rho$ is the mean dust density along the l.o.s, ${\sigma}_{r}$ is the distance uncertainty, and ${\sigma}_{{a}_{d}}$ is the uncertainty in attenuation propagated from the distance uncertainty. The total input attenuation uncertainty in the model (${\sigma}_{{a}_{tot}}$) is then
\begin{eqnarray}
{\sigma}_{{a}_{tot}} = {({{\sigma}_{{a}_{d}}}^{2} + {{\sigma}_{a}}^{2})}^{1/2}
\end{eqnarray}
where ${\sigma}_{a}$ is the measured attenuation uncertainty.\\
This provides us with an upper limit on the input attenuation uncertainty as a result of the distance uncertainty: if a star is in or close to a high density environment, then varying its distance would cause a significant difference in the attenuation measurements; otherwise, if changing the star's location would not impact its attenuation measurement, this would overestimate its attenuation uncertainty.
We use this maximum possible uncertainty as the input to our model to account for both the extinction and distance uncertainties. This provides, in the higher-density regimes, more flexibility for the model to capture the underlying dust density variations when considering the neighbouring correlations.

Another major improvement we report in this paper is on the computational limitation. This was an issue we discussed in \cite{Rezaei_Kh_17} and \cite{Rezaei_Kh_18b}: since we consider the correlation between all points in space, the further the stars, the more cells we have, resulting in a more computationally expensive calculation. We have overcome this problem by partitioning our dataset into different slices as we now explain:\\
First, we infer the dust density for an inner sphere with radius of one correlation length, ${\lambda}$, using stars within two correlation lengths. We then treat the inferred inner densities as known values, taking into account their correlated inferred density uncertainties, to derive the densities for the next layer and continue until we reach our desired distance. From the second layer on, points in different directions could be more distant than ${\lambda}$ from the l.o.s. So to predict dust densities along a given l.o.s, we only consider stars that lie within the correlation length of that l.o.s. This corresponds to a cylinder with radius equal to ${\lambda}$ (extending from the sphere of radius ${\lambda}$ that was the first layer). This approach ensures that we include all relevant neighbouring stars for the correlation computations, yet speeds these by ignoring stars beyond a distance ${\lambda}$.

Using this setup we can predict the dust densities for any point in space. The resolution of the final map is set by the density of the input data, which is on average equal to the typical separation between the input stars.


\section{Input Data}\label{sec:data}

\begin{figure*}
\resizebox{\hsize}{!}{\includegraphics[clip=true]{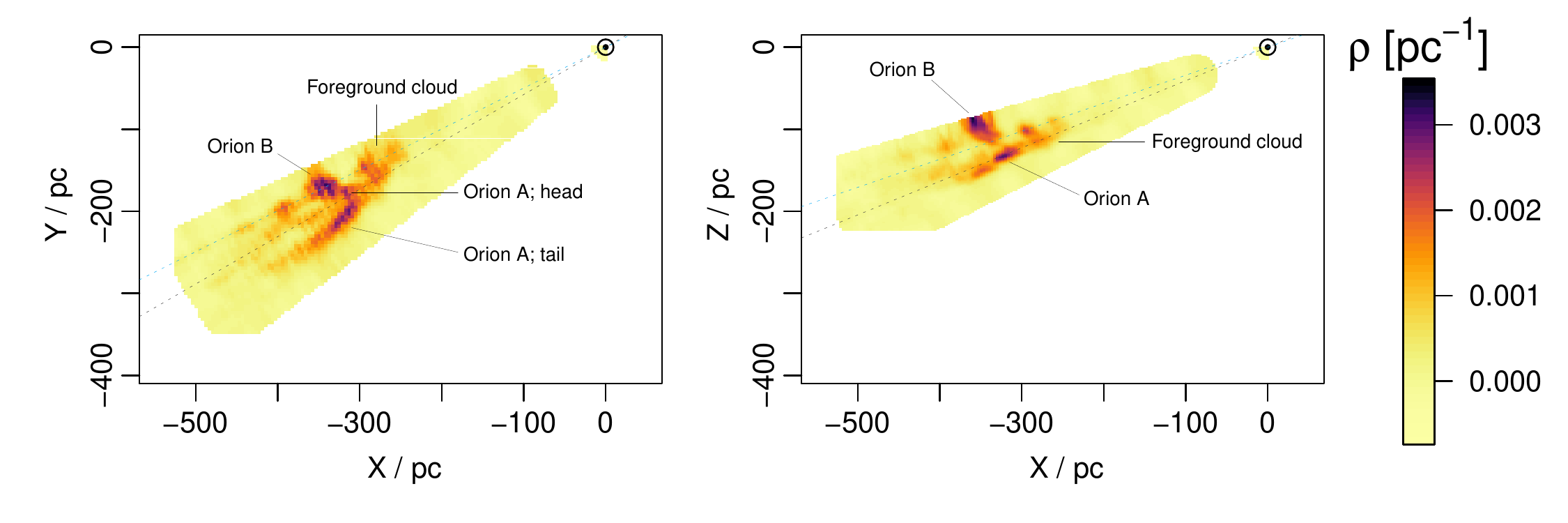}}
\caption{Two Cartesian projections of the 3D dust distributions in Orion. The Sun is at (X, Y, Z) = (0 , 0, 0), with X increasing towards the Galactic centre and Z points to the North Galactic pole,
perpendicular to the Galactic disk. The left panel looks through the Galactic plane from north to south and the right panel is perpendicular to that of the left, having the Galactic height as the vertical axis. The presence of the foreground bubble structure is evident in both projections. Also the extent of the tail of Orion\,A to large distances is clearly seen from the left panel. The dashed lines are two l.o.s passing through different parts of the foreground cloud analysed further in Fig \ref{fig:los}.
The predictions are made on regular grids for every 0.5 degrees in the Galactic $\emph{l}$ and $\emph{b}$, and every 10 pc in distance. The 2D image is then produced by applying a smoothing kernel (with 4-pc scale length) to handle the missing pixels. In order to not produce extra smoothing than that of the method, the length scale of the kernel is chosen to be relatively small; hence, the distance gridding is still apparent in the left panel. \label{fig:3d}}
\end{figure*}

We use data from the second Gaia data release \citep[Gaia DR2,][]{Gaia18a} to get the 3D positions of stars. It is important to note that Gaia provides parallaxes for stars and not distances and since we do not cut on the precise parallax measurements, we need to account for the noisy parallaxes to get the distance information. We therefore use the catalogue of \cite{Bailer-Jones18} who infer distances to Gaia sources from noisy parallaxes. For the uncertainty in the estimated distance, we take the average of the estimated lower and upper confidence interval.

In addition to 3D positions, the method of course needs a measure of extinction towards each star. Gaia DR2 provides extinction measurements for around 88 million sources \citep{Andrae18}. However, due to the strong degeneracy between the extinctions and temperatures of stars, the individual extinctions from Gaia DR2 are less reliable for star-by-star analyses \citep{Andrae18}. We, therefore estimate extinctions using the Rayleigh-Jeans Colour Excess \citep[RJCE,][]{Majewski11} method, which uses near-infrared minus mid-infrared (NIR-MIR) colours of stars in order to get their $K_{s}$-band extinctions. The RJCE method works based on the fact that the distribution of the intrinsic colours of stars in NIR-MIR is so narrow that it can be treated as a known value with some uncertainty. The difference between the observed colours and this intrinsic value is then due to the column of the dust between the star and the observer \citep{Majewski11}. We use H band photometry from the Two Micron All-Sky Survey \citep[2MASS;][]{Skrutskie06} as the near-infrared data and the WISE W2 band photometry \citep{Wright10} as the mid-infrared one. Both catalogues are cross-matched with the Gaia DR2 sources in the Gaia archive\footnote{https://gea.esac.esa.int/archive/}.

In this work, we focus on the Orion\,A region and we select stars within $204^{\circ} < l < 218^{\circ}$ and $-22^{\circ} < b < -15^{\circ}$. As explained in detail in \cite{Rezaei_Kh_18a}, after we derive the extinction values using the RJCE method, we select our final sample based on the positions of the stars on the de-reddened colour-magnitude diagram in order to remove the outliers that happen to get unrealistic high extinction values due to the star being photometrically variable, or so young that it still harbours a dusty disk \citep[see][]{Rezaei_Kh_18a}. This leaves us with about 30\,000 stars out to 700 pc in distance towards this area, making the typical separation between stars to be $\sim 5$ pc. The hyper-parameters of the model set for this dataset (see section \ref{sec:method}) are; ${\lambda = 30 pc}$, ${\theta} = 8 {\times} {10}^{-8}$ ${pc}^{-2}$, and the mean density of the Gaussian process is $1 {\times} {10}^{-4}$ $pc^{-1}$.

\section{Orion dust map} \label{sec:map}

Fig.~\ref{fig:3d} shows 2D Cartesian projections of the inferred 3D dust densities in the Orion\,A region. As can be seen from the figure, there is a bubble-like structure in the foreground of the main Orion\,A filament, at about 350 pc, that expands to larger longitude and latitude compared to Orion\,A. In addition, the extent of the Orion\,A to further distances is clear from the left panel. The head of the Orion\,A, which includes the Orion Nebula Cluster (ONC), appears to be at around 400 pc, while as seen from both projections, the tail of the cloud is extending to distances of about 490 pc, making the total length of the Orion\,A cloud to be over 100 pc.

\begin{figure*}[ht!]
\resizebox{\hsize}{!}{\includegraphics[clip=true]{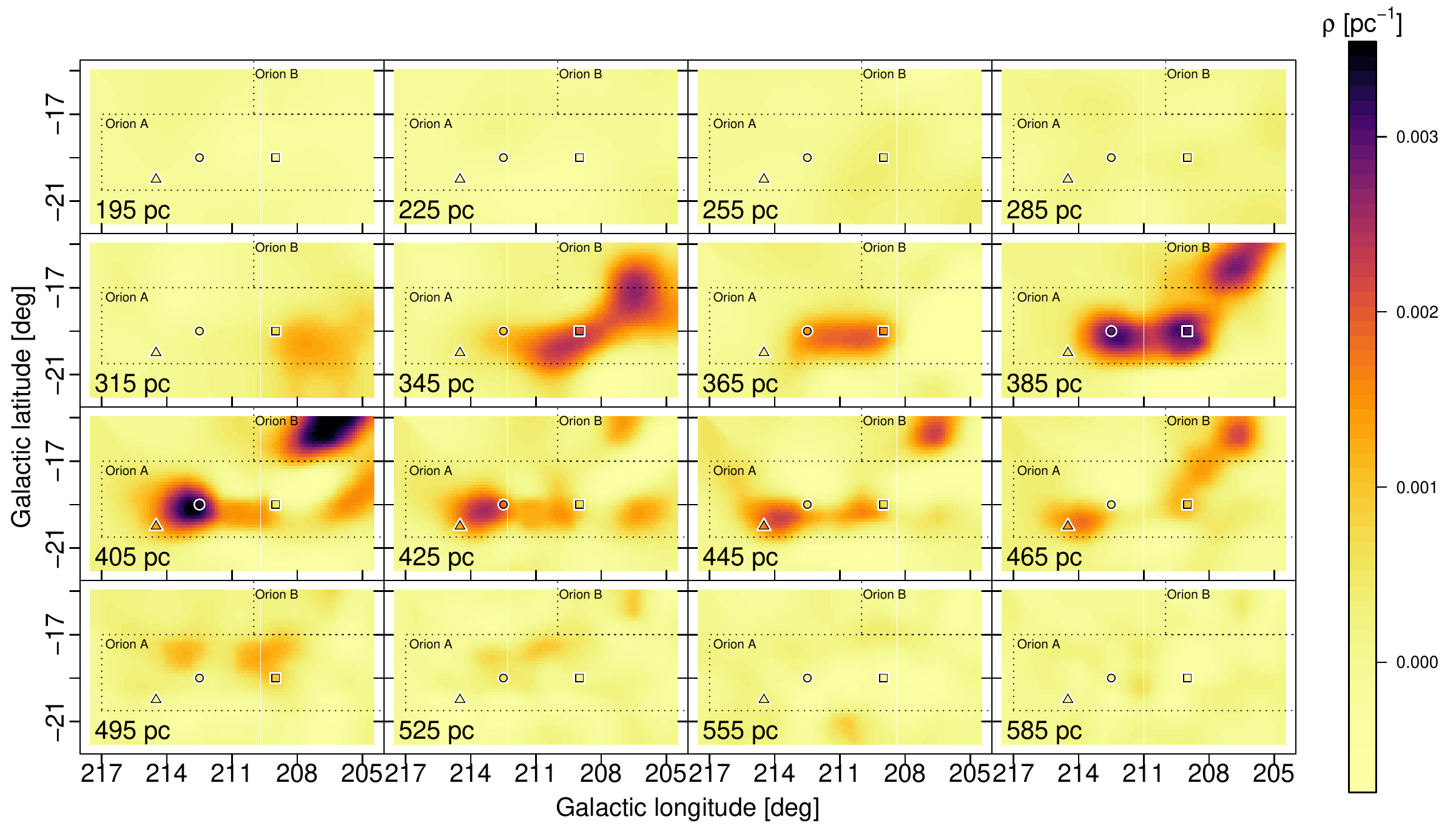}}
\caption{Dust density predictions on the plane of sky. Each panel represents a slice through the cloud at fixed distances (every 30 pc at closer distances and 20 pc around the main structure). The presence of the foreground cloud to Orion\,A is evident at the 345-pc panel. The tail of the Orion\,A (around $l=213$, $b=-19$) appears in multiple panels, illustrating the extent of the cloud to larger distances. The over-density at about 400 pc and higher latitude belongs to the lower part of the Orion\,B. The boundaries of Orion\,A and Orion\,B in the sky projection are shown by dotted lines \citep{Lombardi11}. The three symbols (square, circle and triangle) represent specific l.o.s along the Orion\,A filament investigated in Fig.~\ref{fig:3_los}. For illustration purposes, the image is smoothed with the scale length of 0.4 degrees (see Fig.~\ref{fig:3d} for more detail of the plotting method).\label{fig:slice}}
\end{figure*}

The distance to and the location of individual parts are better seen in Fig.~\ref{fig:slice} that shows the observer's view of the same region. Each panel indicates a slice through the region at a fixed distance. The foreground structure is clearly seen in the 345-pc panel and seems to be extended upwards to around $l = -16^{\circ}$. The extent of the tail of the Orion\,A is also visible in multiple panels up to further distances.

Our map covers a slightly larger area than that of the Orion\,A; consequently, the lower part of the Orion\,B (higher densities above latitude of $-17^{\circ}$) appears in our results. As shown in figures \ref{fig:3d} and \ref{fig:slice}, apart from the main Orion\,B over-density at slightly above 400 pc, a background component is revealed at the distance of around 460 pc. A dedicated study on the Orion\,B cloud will be carried out in a future work.

\begin{figure}[ht!]
\begin{center}
\includegraphics[width=0.50\textwidth, angle=0]{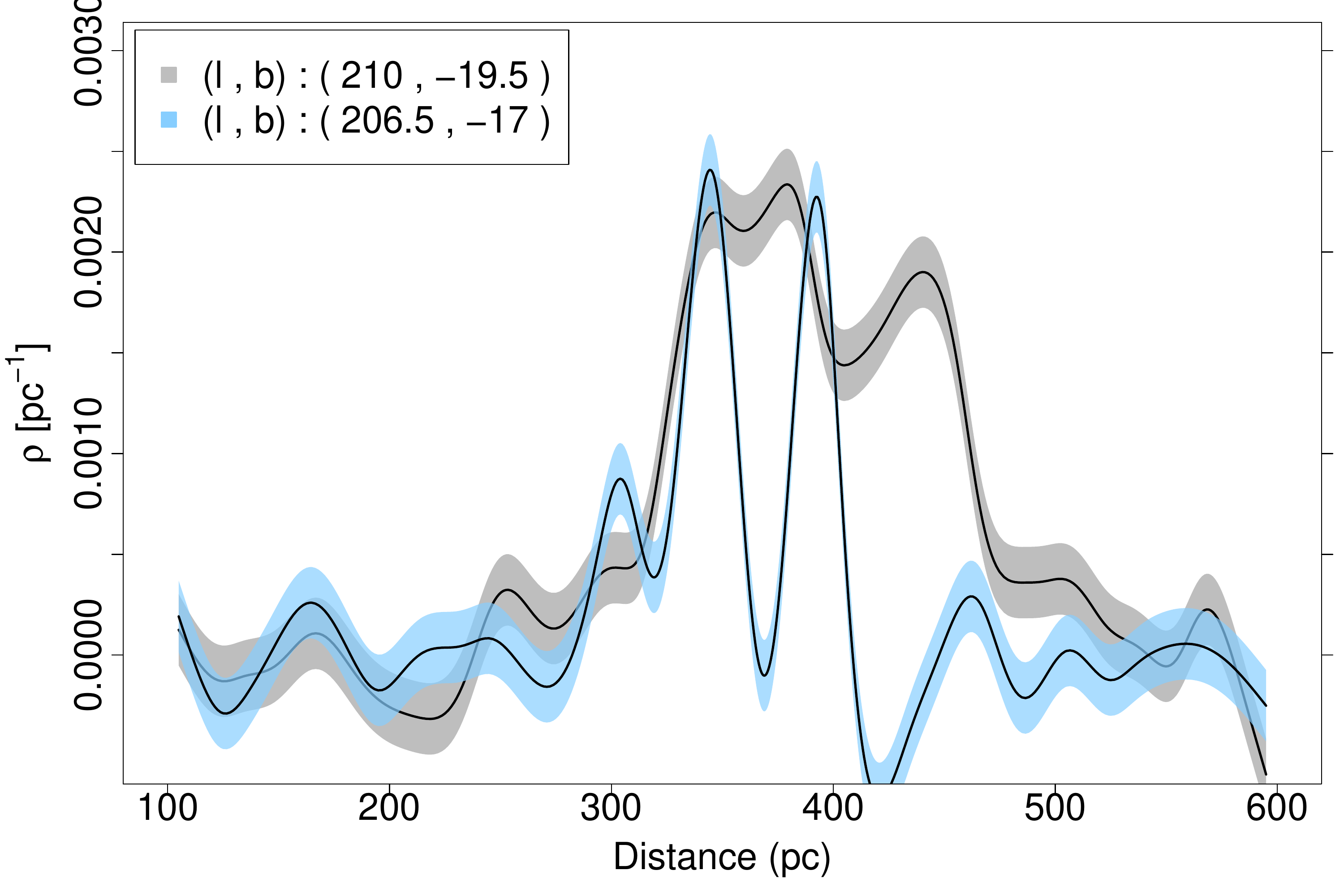}
\caption{Dust density vs. distance for two different l.o.s. towards upper (blue shades) and lower (grey shades) parts of the foreground structure (over-plotted on Fig.~\ref{fig:3d}). The black line shows the mean and the shades represent one standard deviation (also computed by the Gaussian Process model). The grey-shaded first peak represents the foreground cloud while it seems to be connected to the main cloud in the background (the second peak). The two blue peaks illustrate the front and back edges of the foreground cloud. \label{fig:los}}
\end{center}
\end{figure}

\begin{figure}[ht!]
\begin{center}
\includegraphics[width=0.50\textwidth, angle=0]{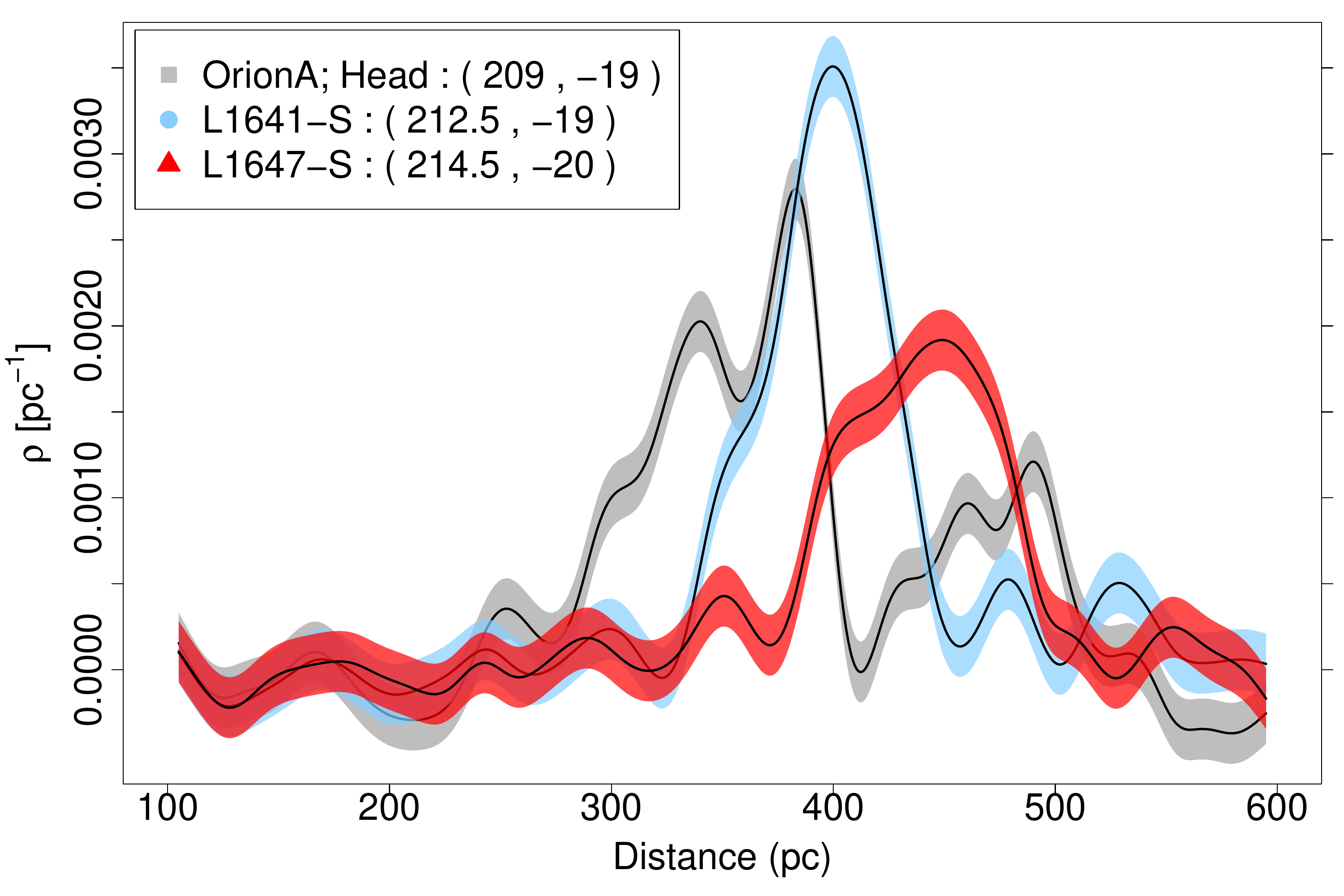}
\caption{Same as Fig.~\ref{fig:los} but for three other l.o.s; one towards the Orion\,A head (in grey), another one towards the middle of the filament, at the location of the L1641-S cluster (in blue), and the third one towards the tail of the Orion\,A and at the location of the L1647-S cluster (in red). The projected sight lines are over-plotted on Fig.~\ref{fig:slice} as a reference. \label{fig:3_los}}
\end{center}
\end{figure}

\begin{figure*}
\resizebox{\hsize}{!}{\includegraphics[clip=true]{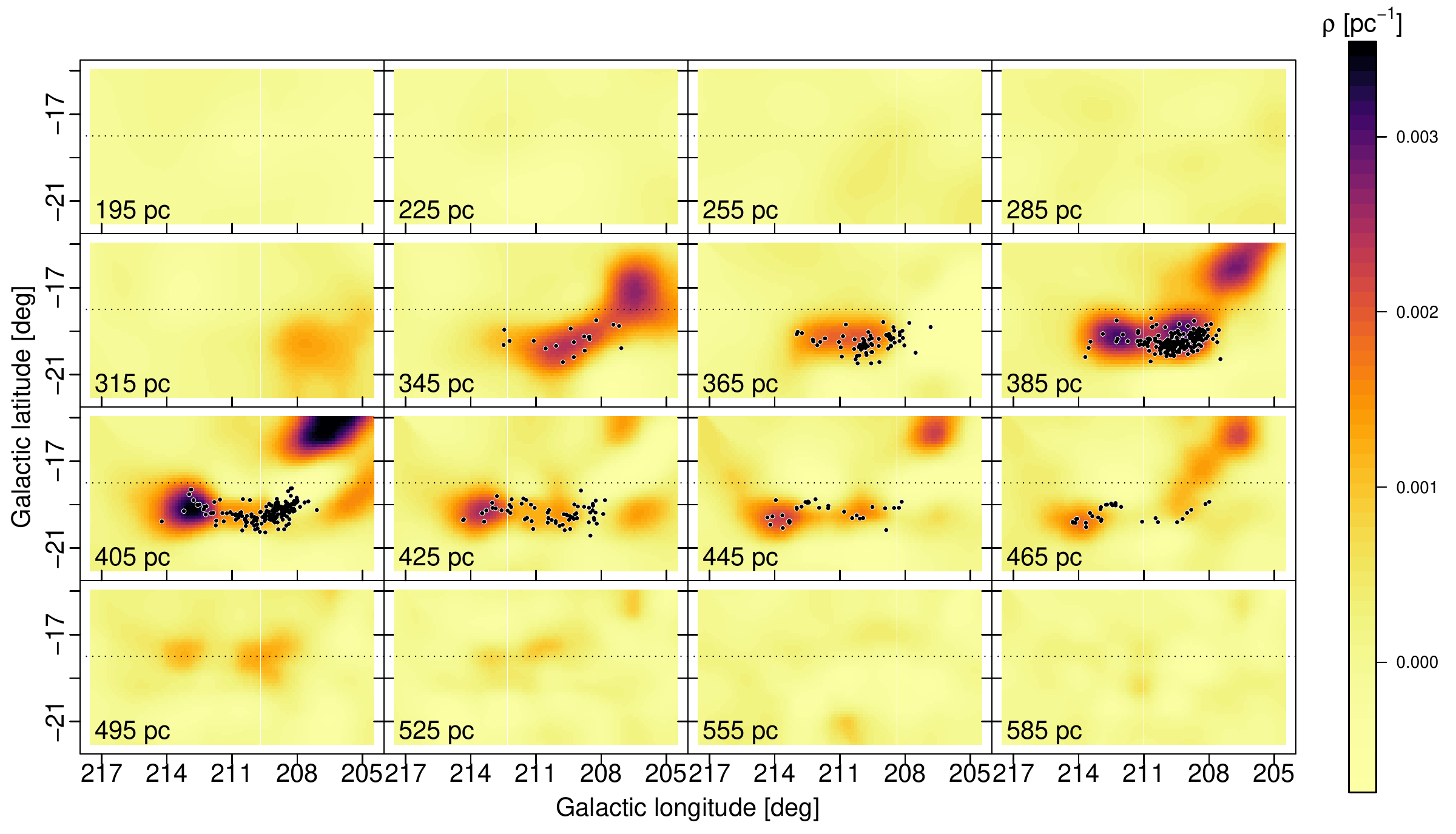}}
\caption{Same as Fig.~\ref{fig:slice} with YSOs of \cite{Grossscheld18} over-plotted as black circles. For each distance panel, we selected YSOs within 5 pc of the slice. The YSOs follow our dust pattern quite well. The dotted line indicates the limit of the YSO survey in latitude. \label{fig:image_YSOs}}
\end{figure*}

In the previous figures we plotted only the mean of our density predictions, while our method provides the probability density distribution for each point. To investigate the uncertainties of the predictions and the significance of the predicted dust clouds we look at two lines-of-sight towards the foreground cloud and plot the predicted dust densities and their uncertainties as a function of distance, as illustrated in Fig.~\ref{fig:los}. The two l.o.s are also shown on the 3D projections in Fig.~\ref{fig:3d}. The presence of the foreground cloud is evident in both lines-of-sight as significant over-densities. As demonstrated in the plot, the lower part of the foreground structure (grey) is connected to the main Orion\,A filament at larger distances, while the density of the upper part (blue) drops significantly after the foreground cloud, indicating the separation between the two at the upper part.

It is important to note that our model infers densities for ``given" points in space, i.e. the location of the predicting points are fixed. As a result, the model by design does not provide distance uncertainty for the predictions. The model does, however, provide uncertainties for the inferred densities. Similar to the way we accounted for the distance uncertainty in the input data (see eq. \ref{eq:err}), inferred density uncertainties can be translated into distance uncertainties. This way the typical distance uncertainty is estimated to be 10 pc in our map.

\section{Discussion} \label{sec:discussion}

\subsection{The Orion\,A structure and young stellar populations}

\begin{figure}[ht!]
\begin{center}
\includegraphics[width=0.50\textwidth, angle=0]{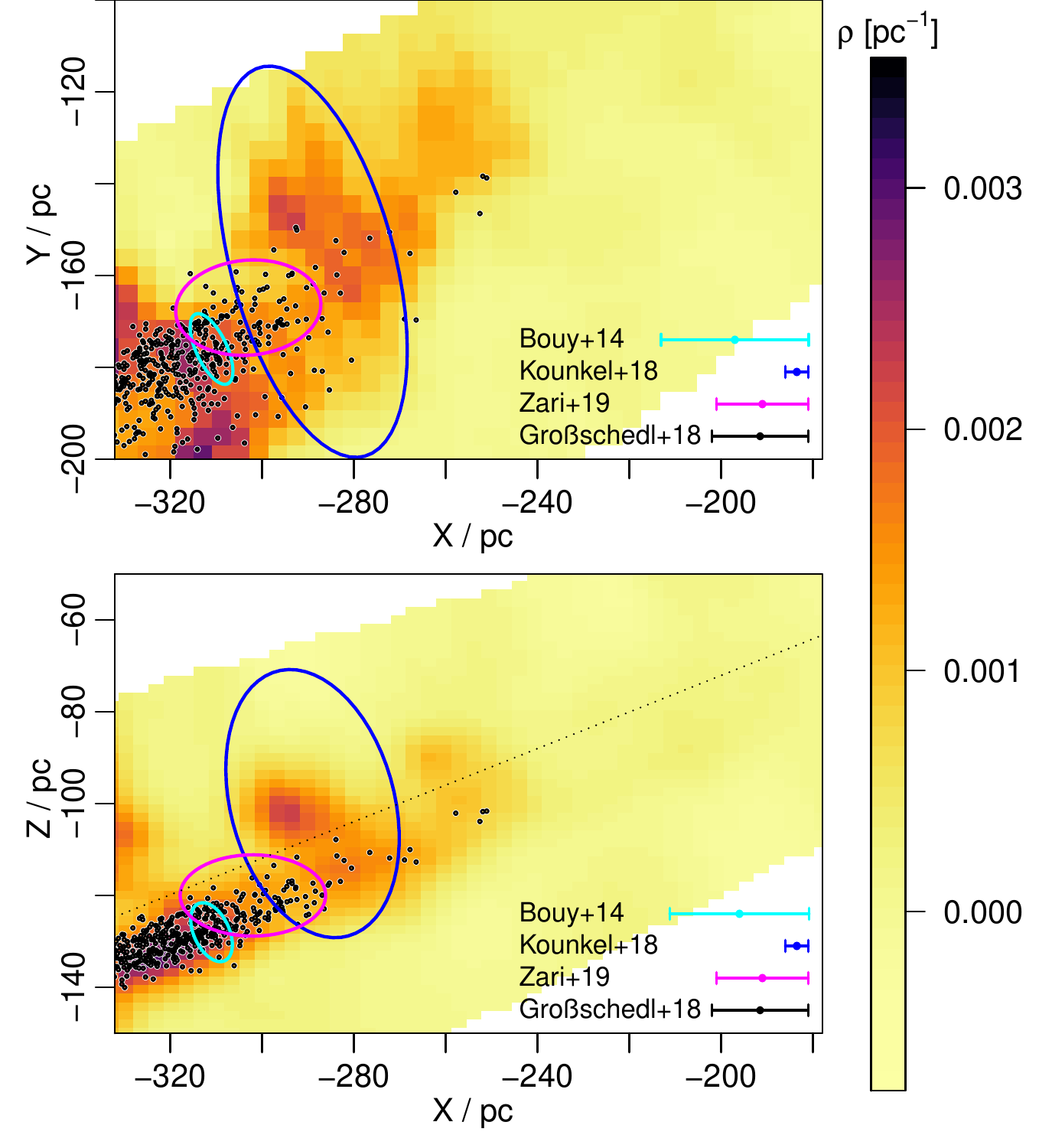}
\caption{Same as Fig.~\ref{fig:3d} zoomed in to the location of the foreground cloud, with stellar associations over-plotted. The cyan, magenta, and blue ellipses represent the location of stellar populations from \cite{Bouy14}, \cite{Zari19}, and \cite{Kounkel18} respectively. The black dots are YSOs from \cite{Grossscheld18}. Each group has stars with different l, b, and distance that makes them appear within their corresponding ellipse on these plots, therefore, the size of each ellipse shows the distribution of stars in the 3D space. The error bars on the bottom right corner of the panels
illustrate the error on the centre position for the groups shown by ellipses, and the typical uncertainty for the individual stars in \cite{Grossscheld18}. The dotted line in the lower panel represents the limiting latitude of the YSO survey. \label{fig:stellargroups} }
\end{center}
\end{figure}

The foreground bubble in our map has not been reported prior to this work. This could be mainly due to the projection effects, lack of distance resolution, or the limitations of the underlying techniques. From about 30,000 stars in our sample, nearly 5,000 of them are within 300 -- 400 pc distance, of which $\sim2200$ are located around the coordinates where we discovered the foreground cloud. This indicates the high number statistics involved in our analysis.

To elaborate this more, we compare the recent results of \cite{Grossscheld18} with ours. They use the 3D positions of about 700 YSOs in the vicinity of the Orion\,A filament to find the average distances of stars in 1-degree longitude bins. They have reported that the Orion\,A cloud is not a straight filament as is seen in 2D projection, but has an elongated tail extending to larger distances \citep{Grossscheld18}. The tail of Orion\,A is clearly seen in the left panel of Fig.~\ref{fig:3d} in our work, here we focus on three specific l.o.s as probed by \cite{Grossscheld18} which we show in Fig.~\ref{fig:3_los}: one towards the head of the Orion\,A, and the other two towards the two star clusters L1641-S and L1647-S. Going from the head to the L1641-S and L1647-S, the peaks of the density distribution moves from smaller to larger distances, confirming the results of \cite{Grossscheld18}. We note that the l.o.s towards the head of the Orion\,A shows two peaks, the closer one being the foreground cloud (as was discussed in the previous figures/sections). We also see that the dust density towards the tail of the cloud shows a wide distribution meaning that the tail starts as close as about 400 pc, then extends to further distances. These points are indeed present in Fig.~3 of \cite{Grossscheld18}: at longitudes of around $208^{\circ}$ to $210^{\circ}$, there are quite a few YSO distances around 350 pc, where we discovered the foreground cloud. In addition, the large scatter in their distance estimates around the middle and tail of the filament is indeed real representing the extent of the cloud. Fig.~\ref{fig:image_YSOs} shows the YSOs in \cite{Grossscheld18}, over-plotted on our distance slice plot. The YSO locations nicely match our dust distribution in the regions covered by \cite{Grossscheld18}; both at the location of the foreground cloud, and along the main Orion\,A filament.

The location of our foreground structure seems to be consistent with the findings of recent studies focused on the stellar populations towards Orion\,A. \cite{Bouy14} reported a large foreground population towards Orion\,A, loosely clustered around NGC1980 and NGC1981, at a distance of 380 pc. They estimated the age of this population to be between 5 and 10 Myr. \cite{Kounkel18} reported similar results for their Orion D structure where, especially around NGC1981, stars are found at a distance of 357 $\pm$3\,pc. \cite{Jerabkova19} also confirmed the presence of multiple stellar populations towards the ONC and indicated that the older population is closer to us than the ONC. \cite{Zari19} confirmed that the Orion OB association consists of numerous groups with different ages and kinematics and suggested a complex star formation history. In particular, their B1 group (distance of 365 $\pm$10\,pc, age of 10$\pm$0.5\,Myr) appears to be located around our reported foreground dust structure. Figure \ref{fig:stellargroups} illustrates the location of the aforementioned stellar populations over-plotted on our predicted foreground cloud. The fact that the location of our foreground dust cloud and that of older stellar populations roughly coincide, suggests that star formation in the Orion region might indeed have started at a closer distance to the Sun than that of the ONC.  Supernova explosions and stellar winds from massive stars could have triggered a new episode of star formation in the ONC and other younger clusters in the region. Our foreground dust cloud could then be a remnant of a previous star formation episode.

\subsection{Reconstructed extinction map}\label{sec:ext}
\begin{figure}[ht!]
\begin{center}
\includegraphics[width=0.50\textwidth, angle=0]{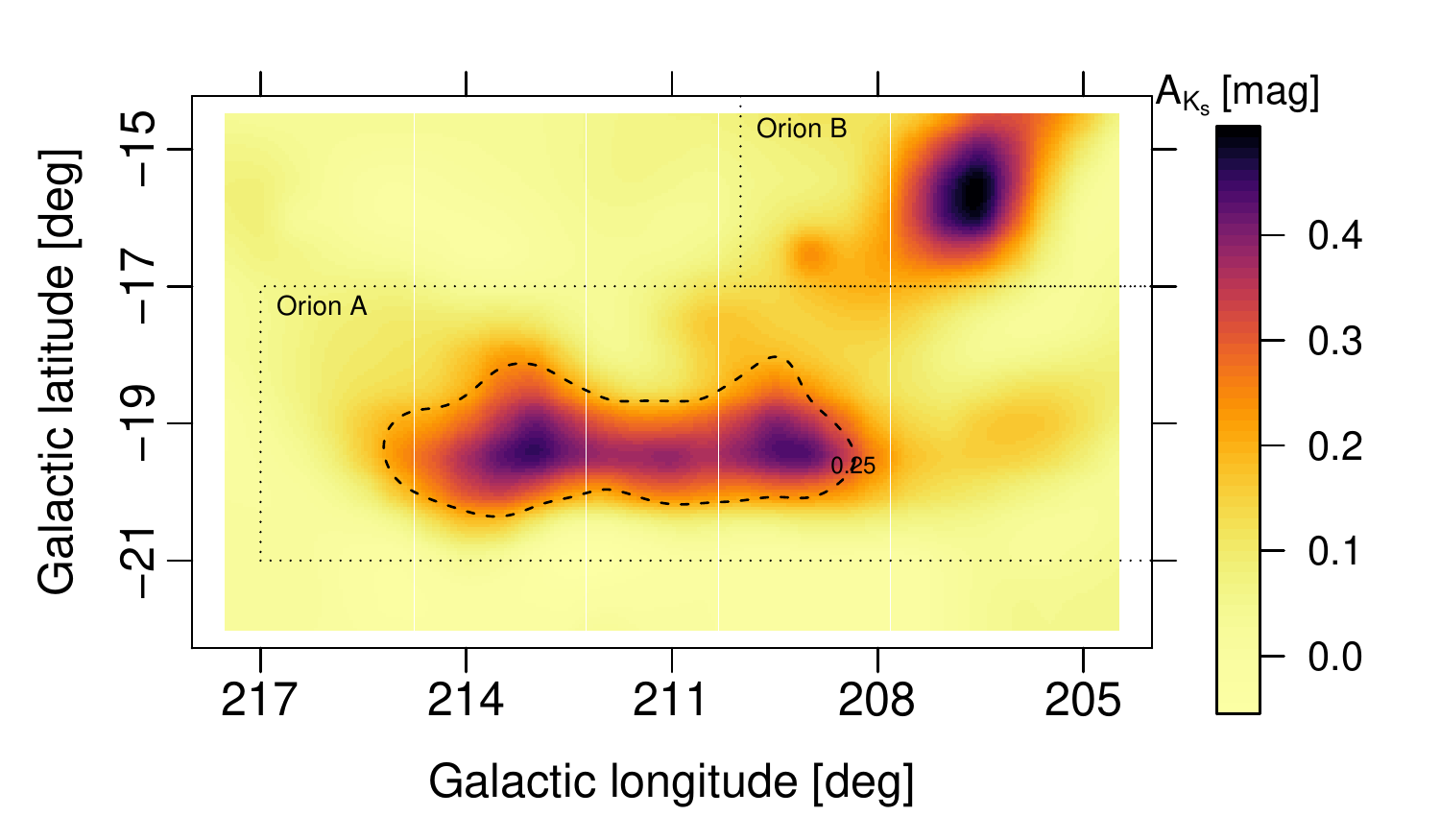}
\caption{Reconstructed extinctions up to 600 pc using the predicted dust densities. The plot is made by summing up the predicted densities in each distance grid along each l.o.s, multiplied by the grid size (10 pc here). The Orion\,A filament and part of the Orion\,B is nicely recovered by the predictions. The dashed contour represents 0.25 mag extinction threshold from \cite{Green19} extinction map, smoothed to the same resolution as our map. The boundaries of Orion\,A and Orion\,B are marked by dotted lines from \cite{Lombardi11} extinction map as a reference too. For illustration purposes the image is smoothed as explained in Fig.~\ref{fig:3d}. \label{fig:re_ext}}
\end{center}
\end{figure}

We can test our inferred 3D distribution of the dust in this region by recreating a 2D extinction map. We do so by summing up the predicted densities along each l.o.s. from the Sun to 600 pc, as presented in Fig.~\ref{fig:re_ext}. We superimpose the borders of Orion\,A and Orion\,B from the \cite{Lombardi11} extinction map. The main Orion\,A structure can be seen in the central part, together with the lower part of the Orion\,B that appears at the top of the plot. Around the longitude of $214^{\circ}$, a ring-type structure is visible. This is also seen in the dust column density map from {\it Planck} (see Fig.~\ref{fig:Orion_PlanckBandNH}). The ring was thought to be associated with the $\kappa$-Ori star located at the centre of the ring in the 2D projection. The distance to the $\kappa$-Ori star is estimated to be 200 $\pm10$ pc from Hipparcos \citep{vanLeeuwen07}; however, there does not seem to be any associated over-density in our map at those distances. Since the star is fairly bright (V $\sim$ 2 mag), Gaia has not yet published its parallax. We do not find a ring structure in our 3D map that would correspond to what is seen in the 2D projection. More focused study on the $\kappa$-Ori region will be carried out in future work.

It is important to note that we use optical data from Gaia, which omits stars in very dense parts of the cloud due to the high obscuration by dust. When computing the correlations, all stars within a correlation length contribute to the inference of the density for a point in that volume. Since lower extinction stars are represented more in the input data, the average extinction in dense regions is less than it would be in the presence of deeper dataset. Therefore, the values we report here for extinctions in the higher density regime (like around the ONC) could be an underestimate of the total extinction.
To elaborate this, we compare our reconstructed extinctions towards Orion\,A with the extinction map of \cite{Green19}, which is primarily based on optical and near-infrared photometry from Pan-STARRS 1, combined with near-infrared photometry from 2MASS, and Gaia parallaxes. We smooth their map to the resolution of our reconstructed extinction map to have a reasonable comparison. The contour in Fig. \ref{fig:re_ext} represents 0.25 mag $K_{s}$-band extinction threshold from \cite{Green19}, which indicates a good agreement between the two maps.

\begin{figure}[ht!]
\begin{center}
\includegraphics[width=0.50\textwidth, angle=0]{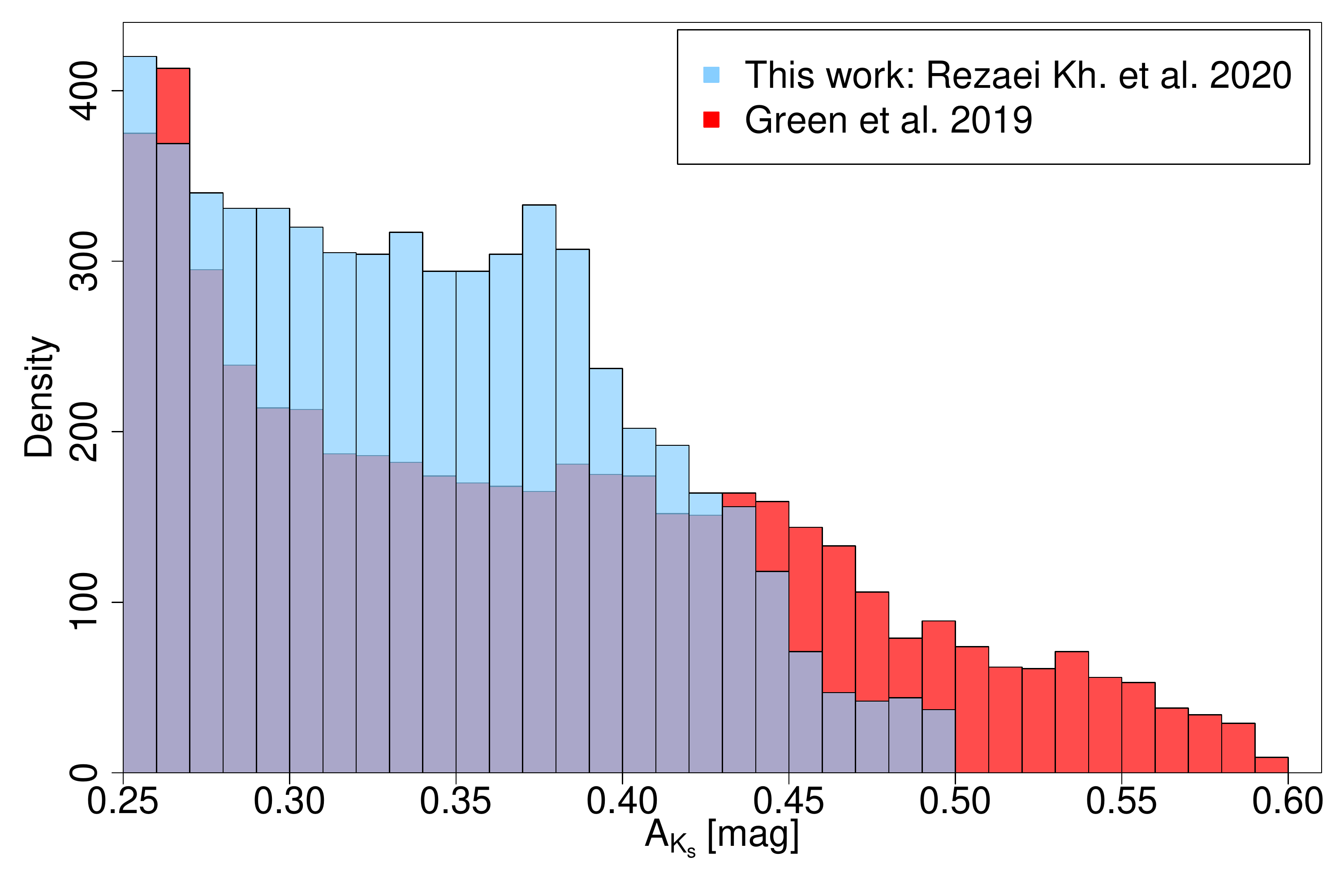}
\caption{Histograms of the extinction values above $A_{K_{s}} = 0.25$ mag for our reconstructed extinction map (blue) and the map of \cite{Green19} (red). The latter map is smoothed to the resolution of our map in Fig. \ref{fig:re_ext}. Where the two histograms overlap, the colour appears in purple. \label{fig:hist}}
\end{center}
\end{figure}

\begin{figure}[ht!]
\begin{center}
\includegraphics[width=0.50\textwidth, angle=0]{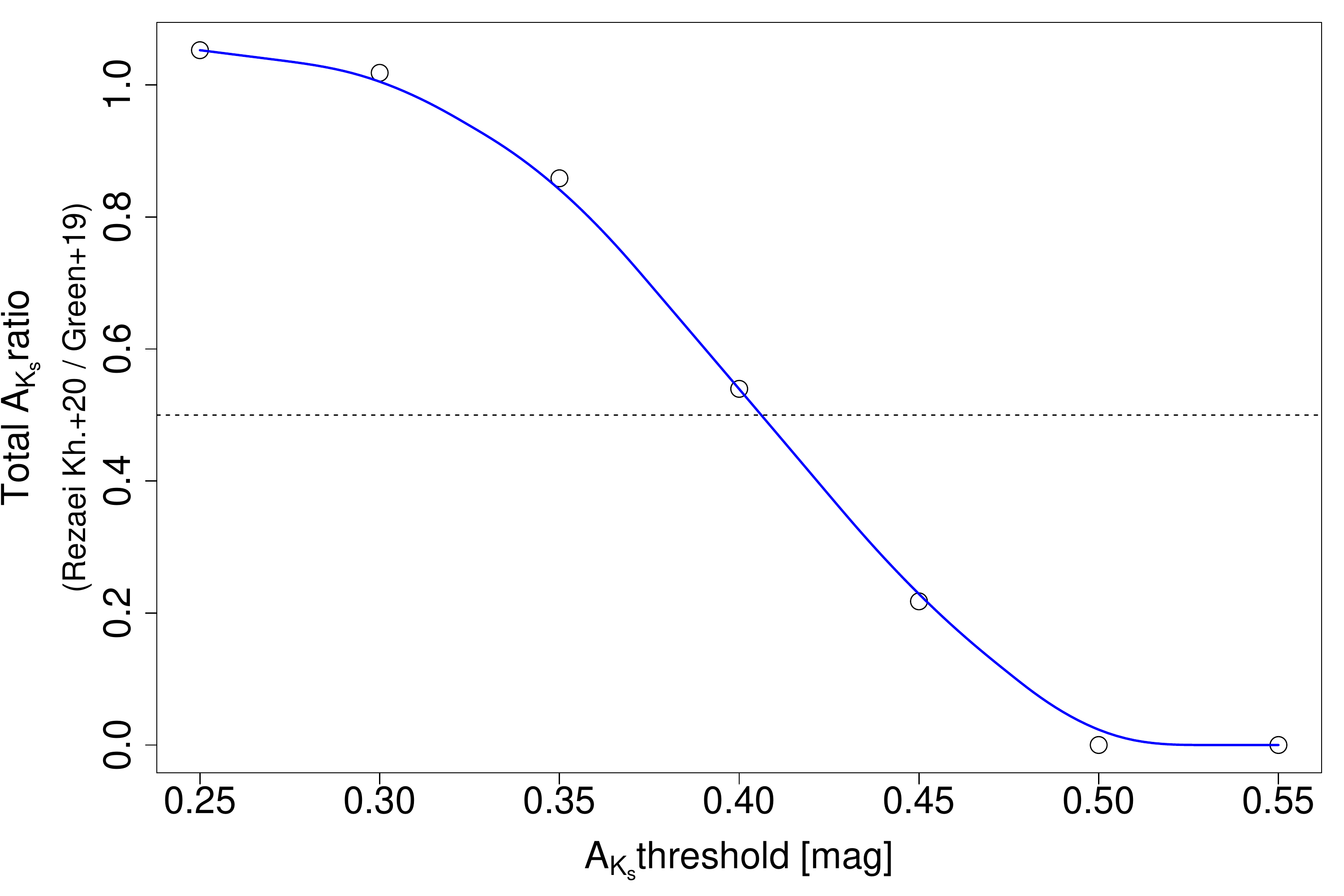}
\caption{Ratio of the total extinctions between our reconstructed extinction map and that of the \cite{Green19} as a function of the extinction threshold used. For each extinction threshold, we calculate the total amount of extinction above that value for each map and plot the ratio between the two (the Y axis). This is a proxy of the total mass of the cloud within that threshold. The blue line is a curve fit to the points using an X-Spline with shape = 0.5 \citep{Blanc98}. The dashed line indicates the total extinction ratio of 0.5. \label{fig:mass}}
\end{center}
\end{figure}

In order to have a more quantitative comparison, we apply 0.25 mag $K_{s}$ band extinction threshold to both maps and compare the two histograms of the extinction distribution above this threshold (Fig. \ref{fig:hist}). While the maximum value of extinction in our map does not reach that of the \cite{Green19}, our map contains more middle-range extinction values ($\sim$ 0.3 - 0.4 mag) than \cite{Green19}, making the total amount of extinction within the threshold almost equal for both maps. This is important because the total amount of extinction is a proxy of the total mass of the cloud. We repeat the same experiment with different extinction thresholds and compare the total extinctions in the two maps. Figure \ref{fig:mass} illustrates the results: for $A_{K_{s}}$ thresholds below 0.35 mag, our reconstructed extinction map recovers almost the same total mass of the cloud as in \cite{Green19}. However, as indicated in the figure, for the extinction threshold of slightly above 0.4 mag, the ratio drops to about half, i.e., our reconstructed extinction map would underestimate the mass of the cloud by 50\% or more if only concentrated on the areas with $A_{K_{s}}$ above $\sim$ 0.4 mag.

\subsection{Magnetic field structure}

\begin{figure}
\centerline{\includegraphics[width=0.5\textwidth,angle=0,origin=c]{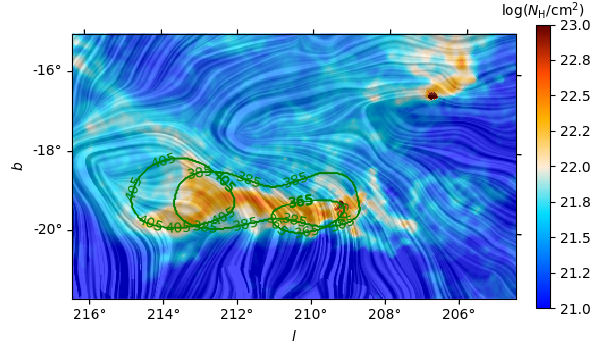}}
\caption{
Dust column density and magnetic fields toward the Orion clouds.
The colours represent the  logarithm of the dust column densities inferred from the {\it Planck} observations \citep{planck2013-p06b}.
The overlaid drapery pattern correspond to the representation of the plane-of-the-sky magnetic field orientation derived from the {\it Planck} polarization observations at 353\,GHz obtained using the line integral convolution method \citep[LIC,][]{cabral1993}.
The green contours correspond to the areas defined by the extinction threshold $A_{\rm K_{s}}$\,$\geq$\,0.06 mag in the extinction slabs at the indicated distances.
}
\label{fig:Orion_PlanckBandNH}
\end{figure}

One application of our study is the analysis of the plane-of-the-sky magnetic field orientation ($\psi$) revealed by the polarized emission from interstellar dust grains at submillimeter wavelengths \citep[see,][ for a review]{andersson2015}.
The magnetic field orientation inferred from the observations by {\it Planck} at 353\,GHz toward Orion \citep{planck2014-XXI}, shown in Fig.~\ref{fig:Orion_PlanckBandNH}, reveals a variety of bends and kinks that can be attributed to the combined effect of magnetic tension, gravity, and turbulence \citep[see for example,][]{planck2014-XX,planck2015-XXXV}.
However, it is also possible that some of the features in the map are the sum of the contributions from the magnetic fields in dust over-densities located at different distances, which are overlapped in projection.

The observed polarization can be modelled as a series of dust slabs located at different distances, each with its own mean magnetic field orientation, like layers in a diorama \citep{martin1974}.
Given that the magnetic field is a vectorial quantity, the contribution from each slab to the total polarization is not proportional to the amount of dust in each slab but depends on the orientation of the magnetic field.
We evaluate here whether the slab model can provide further information on the structure of magnetic field toward Orion\,A by considering the field orientations within dust over-densities at different distances.
To do this, we construct 16 slabs of extinction between 180 and 600\,pc by integrating the dust densities with $\pm$15\,pc of the distances shown in Fig.~\ref{fig:image_YSOs}.
For each distance slab, we identify the region in which the extinction is above a specified threshold and calculate the mean orientation of the magnetic field $\left<\psi\right>$ within that area.
The changes in $\left<\psi\right>$ in different slabs indicates that there are particular dust over-densities that might be responsible for the observed field orientation. If the values of $\left<\psi\right>$ are roughly constant across distances, this method is insufficient to relate the field orientations to the dust structure in 3D.
\begin{figure}[ht!]
\centerline{\includegraphics[width=0.5\textwidth,angle=0,origin=c]{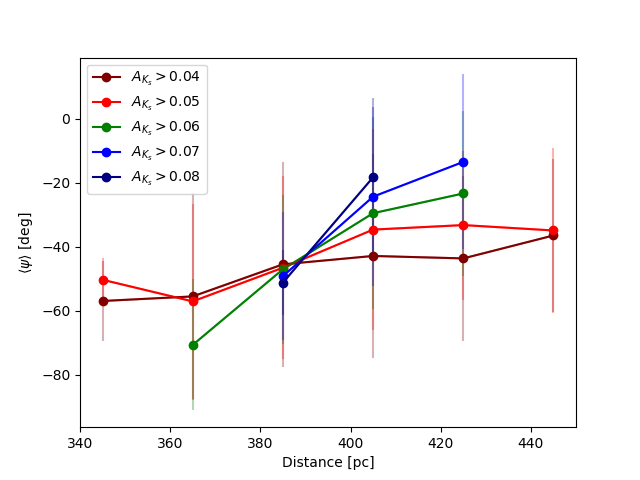}}
\caption{
Mean orientation angles $\left<\psi\right>$ of the plane-of-the-sky magnetic field estimated from the {\it Planck} 353-GHz observations within the regions defined by the indicated extinction thresholds in the 30-pc extinction slabs centred on the values indicated in the $x$-axis.
}
\label{fig:meanPsiVSdistance}
\end{figure}

Figure~\ref{fig:meanPsiVSdistance} shows the values of $\left<\psi\right>$ in the regions defined by five different extinction thresholds between $A_{\rm K_{s}}$\,$=$\,0.04 and 0.08 mag in the slabs at six different distances.
We see that for the extinction thresholds $A_{\rm K_{s}}$\,$<$\,0.06 there is no significant variation in $\left<\psi\right>$, most likely because the magnetic field orientation is being averaged over similarly large areas on the plane of sky for all distance slabs resulting in similar values for $\left<\psi\right>$.
However, for the extinction thresholds $A_{\rm K_{s}}$\,$\geq$\,0.06 we see changes of up to 50$^\circ$ within the regions defined by the extinction thresholds at different distances.

\begin{figure}[ht!]
\centerline{\includegraphics[width=0.5\textwidth,angle=0,origin=c]{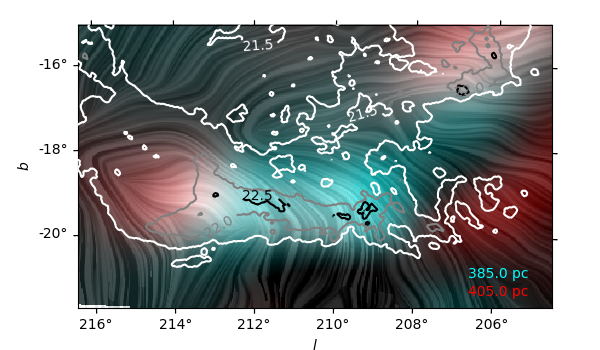}}
\caption{
Extinction in the 30-pc slabs around the distances 385 and 405\,pc, shown in red and cyan as a two-colour image.
Overlap in the extinction in the two slabs is show by the brighter colour between cyan and red (e.g. around $l = 213^{\circ}, b = -19^{\circ}$).
The contours correspond to the logarithm of the dust column densities inferred from the {\it Planck} observations, $\log_{10} (N_{\rm H}/{\rm cm})^{-2}$.
The drapery pattern corresponds to the orientation of the plane-of-the-sky magnetic field orientation inferred from the {\it Planck} 353-GHz polarization observations.
}
\label{fig:distanceRGBandBfield}
\end{figure}

The contours in Fig.~\ref{fig:Orion_PlanckBandNH} illustrate an example of regions where the most significant changes in $\left<\psi\right>$ are found for the particular case $A_{\rm K_{s}}$\,$\geq$\,0.06.
The figure clearly shows that the mostly vertical magnetic field orientation at $l = 209^{\circ}, b = -20^{\circ}$ is associated with the dust over-densities in the slabs centred around 385\,pc. In contrast, the mostly horizontal magnetic field at $l = 214^{\circ}, b = -19^{\circ}$ corresponds to the over-density in the slab corresponding to 405\,pc, as illustrated in Fig.~\ref{fig:distanceRGBandBfield}.
The region of overlap between the 385-pc and the 405-pc thresholds shows a field orientation at around 45$^\circ$. This could be the result of the superpositions of the orientations in the two slabs, although the current analysis cannot provide further evidence of that scenario.

The fact that different dust components along a l.o.s are causing changes to the magnetic field orientation is important for three reasons.
First, it indicates that the models of the magnetic field orientation, such as those presented in \cite{heiles1997} and \cite{tahani2019}, are limited when considering Orion\,A as just one object. They would instead need to account for the fact that the field is changing its orientation in a dust structure that is just 30\,pc behind the main body of the cloud.
Second, it indicates that the field is bending around and within the cloud. This suggests that this structure is not fully dominated by one single 10-pc-scale magnetic field but rather is showing the combined effects of trans-Alfv\'{e}nic turbulence (i.e. kinetic and magnetic energy densities are comparable) and gravitational collapse \citep[][]{hennebelleANDinutsuka2019,pattle2019}.
Third, it provides an example of a technique that can complement and enhance the studies of the magnetic field in 3D using polarized starlight observations \citep[see for example][]{panopoulou2019}.

\section{Concluding remarks} \label{sec:conclusion}

We have developed a sophisticated mapping technique that takes into account the neighbouring correlation in the space using a Gaussian process prior with a truncated covariance function. The method also considers the distance and extinction uncertainties to individual stars, enabling us to provide a detailed 3D dust map of local molecular clouds. We have presented the results towards the Orion\,A region where, for the first time, we reported a bubble-like dust over-density at 350 pc in front of the previously known Orion\,A cloud. We also illustrated the whole shape of the cloud with its extended tail to distances of about 490 pc. This indicates a length of more than 100 pc for the filament. We also reported a background component to Orion\,B at a distance of 460 pc. 

The presence of stellar associations older than that of the ONC around the same location as our foreground over-density suggests an early star formation episode in front of the ONC. This could have triggered subsequent episodes of star formation in the region.

We also analysed the magnetic field orientation in the plane of the sky. We connected variations in the magnetic field orientation angles with variations in dust density along the l.o.s. This can provide valuable information for understanding the magnetic field distribution in 3D around a star-forming region, which is a crucial step forward in our understanding of the role of the magnetic field in the process of star formation.

\section*{Acknowledgments}
We would like to thank the referee for the very supportive and thoughtful comments that helped to improve the paper. This work has made use of data from the European Space Agency (ESA)
mission {\it Gaia} (\url{https://www.cosmos.esa.int/gaia}), processed by
the {\it Gaia} Data Processing and Analysis Consortium (DPAC,
\url{https://www.cosmos.esa.int/web/gaia/dpac/consortium}). Funding
for the DPAC has been provided by national institutions, in particular
the institutions participating in the {\it Gaia} Multilateral Agreement. SRK and CBJ acknowledge the Sonderforschungsbereich SFB\,881 ``The Milky Way System'' of the German Research Foundation (DFG) for partially funding this project. JDS acknowledges funding from the European Research Council under the Horizon 2020 Framework Program via the ERC Consolidator Grant CSF-648505.

\bibliographystyle{aa}
\bibliography{Rezaei_Kh._2020_OrionA}

\end{document}